\documentclass[prl,twocolumn,aps,final,floatfix,superscriptaddress,showpacs]{revtex4}
\usepackage{graphicx} 
\usepackage{color} 
\usepackage{amsmath} 
\usepackage[urlcolor=blue, hyperindex, colorlinks, bookmarks=true,linkcolor=black,citecolor=black]{hyperref} 
\usepackage{amssymb}

\newcommand{\dg}{^\dagger}

\newcommand{\bra}[1]{\langle{#1}|}
\newcommand{\ket}[1]{|{#1}\rangle}

\newcommand{\sfrac}[2]{\mbox{$\frac{#1}{#2}$}}

\newcommand{\eqrf}[1]{Eq.~\eqref{#1}}

\newcommand{\sxo}{\hat{\sigma}^x}
\newcommand{\syo}{\hat{\sigma}^y}

\renewcommand{\section}[1]{{\em #1}.---}

\begin{document}

\title{Simple pulses for elimination of leakage in weakly nonlinear qubits}
\date{\today}
\author{F. Motzoi}
\affiliation{Institute for Quantum Computing and Department of Physics and Astronomy, University of Waterloo, Waterloo, Ontario, Canada N2L 3G1}
\author{J. M. Gambetta}
\affiliation{Institute for Quantum Computing and Department of Physics and Astronomy, University of Waterloo, Waterloo, Ontario, Canada N2L 3G1}
\author{P. Rebentrost}
\affiliation{Institute for Quantum Computing and Department of Physics and Astronomy, University of Waterloo, Waterloo, Ontario, Canada N2L 3G1}
\affiliation{Department of Chemistry and Chemical Biology, Harvard University, 12 Oxford St., Cambridge, MA 02138, USA}
\author{F. K. Wilhelm}
\affiliation{Institute for Quantum Computing and Department of Physics and Astronomy, University of Waterloo, Waterloo, Ontario, Canada N2L 3G1}

\begin{abstract}
In realizations of quantum computing, a two-level system (qubit) is often singled out from the many levels of an anharmonic oscillator. 
In these cases, simple qubit control fails on short time scales because of coupling to leakage levels. We provide an easy to implement 
analytic formula that inhibits this leakage from any single-control analog or pixelated pulse. It is based on adding a second control that is proportional to the time-derivative of the first. 
For realistic parameters of superconducting qubits, this strategy reduces the error by an order of magnitude relative to the state of the art, all based on  smooth and feasible pulse shapes. These results show that even weak anharmonicity is sufficient and in general not a limiting factor for implementing quantum gates.
\end{abstract}
\maketitle

Quantum information processing devices are paving new inroads into the understanding of coherent processes and their applicability to high-complexity computation. The fundamental building blocks for these devices are quantum bits (qubits), which are quantum two-level systems. Naturally occurring qubits such as spin $\sfrac{1}{2}$ particles and photons are difficult to both isolate and control. Fortunately, effective qubits can also be manufactured from an anharmonic multi-level system.  These systems can be understood as a generalized atom, and all share the common problem of leakage out of the qubit subspace if the control bandwidth  is comparable to the anharmonicity. Examples of such systems are superconducting qubits [1-15], optical lattices \cite{Maneshi:2008a}, and quantum dots \cite{Schliemann:2008a}. 
 

In superconducting qubits \cite{Clarke:2008a}, researchers have demonstrated single qubit gates \cite{Lucero:2008a,Chow:2008a}, two qubit gates \cite{Yamamoto:2003b,Plantenberg:2007a}, and the use of a quantum bus \cite{Majer:2007a,Sillanpaa:2007a}, suggesting the possibility of a scalable technology. Decoherence has dropped dramatically \cite{Martinis:2005a,Houck:2008a,Bertet:2005a} from the original designs \cite{Nakamura:1999a}, due to both device and operation improvement \cite{Martinis:2005a,Koch:2007a,Vion:2002a}.  To date, the single-qubit accuracy of such systems has been in the range of 98\% \cite{Lucero:2008a}- 99.3\% \cite{Chow:2008a}, which is a major achievement, yet not sufficient for quantum computing.

For large scale computation, gate errors are required below a fault-tolerance threshold, conservatively estimated at $10^{-4}$ \cite{Knill:2005a}. In superconducting qubits such as the Phase \cite{Martinis:2002a,Martinis:2005a,Lucero:2008a} and the Transmon \cite{Koch:2007a,Schreier:2008a,Houck:2008a,Chow:2008a} qubits, long pulses are limited by decoherence and short pulses by excitations out of the qubit space \cite{Chow:2008a,Steffen:2003a}. Thus, one prevalent approach to achieve high fidelity gates is to increase the coherence time; alternatively, the approach we will take is to reduce the gate time by using optimized control pulses that avoid leakage.

Thus far, trying to minimize gate times through better control has been met with limited success. Little improvement has been achieved beyond what is possible with a Gaussian waveform \cite{Bauer:1984a,Steffen:2003a}. Optimal control theory with only amplitude modulation has required pulses which have sharp features \cite{Rebentrost:2008a,Safaei:2008a,Jirari:2005a}. For example, the pulses in \cite{Steffen:2003a,Rebentrost:2008a} require 3 pulse steps and 2 waiting periods, which requires currently impractical pulse shaping hardware.  Furthermore, these pulses are limited to a minimal time, given by the inverse of the anharmonicity. 

We propose to solve this problem using two quadrature controls for effective phase modulation, which allows us to achieve {\em instantaneous} cancellation of leakage using slowly varying controls.  We present an analytical approach to accomplish this called Derivative Removal by Adiabatic Gate (DRAG). Then, we map the full potential of the technique using numerical refinement and consider the case when decoherence is included.

\section{The system}
We consider the lowest three levels of a driven slightly anharmonic energy spectrum with only nearest level coupling; the first two levels comprise the qubit and the third level accounts for leakage, provided its population is minimized during the entire pulse.  That is, in the lab frame, the Hamiltonian can be written as 
\begin{equation}
	H = \hbar\sum_{j=1,2}\left[\omega_{j}\Pi_j + \mathcal{E}(t)\lambda_j(\sigma_j^++\sigma_j^-)\right],
\end{equation} 
where $\Pi_j=\ket{j}\bra{j}$ is the projector for the $j^\mathrm{th}$ level, $\sigma_j^-=\ket{j-1}\bra{j}$ is the lowering operator, and the corresponding transition energies are denoted $\hbar\omega_{j}$, with the ground energy set to zero.  The intrinsic anharmonicity of the system is $\Delta=\omega_{2}-2\omega_{1}$. The parameter $\lambda_j$ weighs the relative strength of the $0-1$  to $1-2$ transition, and without loss of generality we will take $\lambda_1=1$ and $\lambda_2=\lambda$. Driving and control of the system is represented by 
\begin{equation}\label{eq:controls}
\mathcal{E}(t)=\begin{cases}
		 \mathcal{E}^x(t)\cos(\omega_d t)+\mathcal{E}^y(t)\sin(\omega_d t), &\text{$0<t<t_g$ }\\
		0, &\text{otherwise,}
	\end{cases}
\end{equation}
 which is a single frequency carrier with two independent quadrature controls, $\mathcal{E}^x(t)$ and $\mathcal{E}^y(t)$; $t_g$ is the time taken for one gate operation.



Making the rotating wave approximation (RWA) and moving to the interaction frame with respect to the drive frequency $\omega_\mathrm{d}$, the Hamiltonian can be written as
\begin{equation}\label{eq:RotHam}
	H^R = \hbar\sum_{j=1,2}\left[\delta_{j}\Pi_j + \frac{\mathcal{E}^x(t)}{2}\lambda_j\sigma_{j-1,j}^x+ \frac{\mathcal{E}^y(t)}{2}\lambda_j\sigma_{j-1,j}^y\right],
\end{equation}
where $\sigma_{j,k}^x=\ket{k}\bra{j}+\ket{j}\bra{k}$ and $\sigma_{j,k}^y=i\ket{k}\bra{j}-i\ket{j}\bra{k}$. $\delta_{1} = \omega_{1}-\omega_d$ and $\delta_2=\Delta+2\delta_1$ are the detunings of the transitions with respect to the drive frequency $\omega_d$. When the drive frequency is resonant with the qubit frequency ($\delta_1=0$), $\delta_2=\Delta$ is the anharmonicity. 

In the ideal case, $\lambda=0$, the Hamiltonian and its evolution reduce to that of a qubit.  Thus qubit operations, 
\begin{equation}\label{eq:ideal}
	U_\mathrm{ideal} = \mathcal{T}\exp\left[-\frac{i}{\hbar}\int_{0}^{t_g}H^Rdt\right] = e^{i\phi_1}(U_\mathrm{qb}+e^{i\phi_2}\Pi_2),
\end{equation}  can be trivially driven by the controls, Eq. \eqref{eq:controls}, since  the effect is $U_\mathrm{qb}$, which is a unitary that acts only in the qubit subspace, and the global phases $\phi_j$ are irrelevant \cite{Rebentrost:2008a}.
For example, the NOT gate (which without loss of generality we choose to focus on)  is implemented by simply setting $\mathcal{E}^x=\mathcal{E}_\pi$, $\mathcal{E}^y=0$. Then, the time-ordering operator  $\mathcal{T}$ can be dropped and any controls that satisfy $\int_{0}^{t_g}\mathcal{E}_\pi dt=\pi$ 
lead to the desired gate $U_\mathrm{qb} =\sigma_{0,1}^x$.

\begin{figure}[htbp]
\centering
\includegraphics[width=0.4\textwidth]{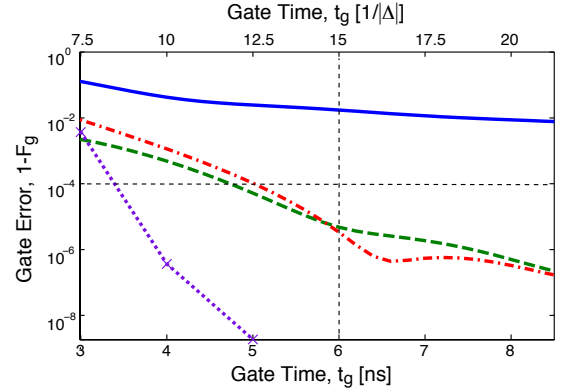}
\caption{Error vs. pulse duration for Gaussian ($\sigma$=$0.5 t_g$) (blue solid), Gaussian ($\sigma$=$0.5 t_g$) with DRAG (red dot-dashed), tangential ($\sigma$=$0.6 t_g$-1.4) with DRAG (green dashed), and GRAPE pulse (purple crosses with dotted guiding line).  The standard Gaussian and DRAG pulses are continuous, while the GRAPE pulse has a pixel size of 1ns. Parameters are $\lambda=\sqrt{2}$, $\Delta=2\pi(-400$ MHz) (typical for the Phase \cite{Martinis:2002a,Martinis:2005a,Lucero:2008a} and Transmon \cite{Koch:2007a,Schreier:2008a,Houck:2008a,Chow:2008a} qubits). The top axis is set to dimensionless units, $1/ |\Delta|$.}
\label{fig:PulseWithoutDecay}
\end{figure}

However, in the more general case when $\lambda$ is non-zero, the pulse can result in large unwanted leakage into the third level.  Fourier analysis can show how this leakage is dictated by the anharmonicity $\Delta$ \cite{Warren:1984a,Steffen:2003a}. The standard approach to diminish this error is to use Gaussian ($\mathcal{E}_G$) \cite{Bauer:1984a} or tangential ($\mathcal{E}_T$) envelope shaping
\begin{equation}\label{eq:gaussian}
\begin{split}
	\mathcal{E}_G(t)=& \left\{A\exp\left[-\frac{(t - t_g/2)^2}{2\sigma^2}\right]-B\right\},\\
	\mathcal{E}_T(t)=& \left\{A\left[\tanh\left(\frac{t}{\sigma}\right)-\tanh\left(\frac{t_g-t}{\sigma}\right)\right]-B\right\},
\end{split}
\end{equation}
where $\sigma$ is the standard deviation for $\mathcal{E}_G$ and the rise-time parameter for $\mathcal{E}_T$. $A$ is chosen such that the correct amount of rotation is implemented (e.g. $\pi$ for a NOT) while $B$ enforces the pulse to start and end at zero. The small frequency response bandwidth of the Gaussian envelope ensures little excitation at the leakage transition frequency. Additionally, Gaussians and tangential pulses have relatively simple pulse-shaping hardware requirements. This being said, these pulses still suffer large errors at short gate times, as shown by the solid blue line in Fig. \ref{fig:PulseWithoutDecay}, where the error is $1\%$ at a gate time of $6$~ns (vertical dashed line),  and scales as $(\mathcal{E}^x_\mathrm{max})^2/\Delta$.  To quantify this error, we use the gate fidelity averaged over all input states existing in the qubit Hilbert space, which using an argument similar to \cite{Bowdrey:2002a} gives 
\begin{equation}
	\begin{split}\label{eq:fid}	
		F_g	=&\frac{1}{6}\sum_{j=\pm x, \pm y, \pm z}\left(\mathrm{Tr}(U_\mathrm{ideal} \rho_j U_\mathrm{ideal}\dg \mathcal{M}_P(\rho_j))\right)
	\end{split}
\end{equation} 
where $\mathcal{M}_P(\rho_j)$ is the actual process in the three dimensionional Hilbert space, $\rho_j$ are the six axial states on the Bloch sphere, and $U_\mathrm{ideal}$ is defined in Eq. \eqref{eq:ideal}.

 
\section{Derivative Removal by Adiabatic Gate (DRAG)} To go beyond these simple pulses we  introduce an adiabatic transformation $V$ that allows us to work entirely in the qubit subspace. This transformation is 
\begin{equation}
V(t)=\exp\left[-i\mathcal{E}^x\left(\syo_{0,1}+\lambda\syo_{1,2} \right)/2\Delta\right],
\end{equation} where ${\mathcal{E}^x}/{\Delta}$ is the adiabaticity parameter.
 For the gates to be identical in both frames we require $V(0)=V(t_g)=\openone$, such that by the end of the pulse it has no net effect. This is achieved by setting $\mathcal{E}^x(0)=\mathcal{E}^x(t_g)=0$. In this frame, to first order in ${\mathcal{E}^x}/{\Delta}$, the effective Hamiltonian $H^V=VH^RV\dg+i\hbar\dot VV\dg$ reads
\begin{equation}
\begin{split}\label{DRAG:Ham}
		&H^V/\hbar \approx \frac{\mathcal{E}^x}{2}\sxo_{0,1} +\frac{\lambda\mathcal{E}_x^2}{8\Delta}\sigma_{0,2}^x + (\delta_2 + \frac{(\lambda^2+2)\mathcal{E}_x^2}{4\Delta})\Pi_2 \\
		&+ (\delta_1 - \frac{(\lambda^2-4)\mathcal{E}_x^2}{4\Delta})\Pi_1 + \left[\frac{\mathcal{E}^y}{2}+\frac{\mathcal{\dot E}^x}{2\Delta}\right](\syo_{0,1} + \lambda\syo_{1,2}).
\end{split}
\end{equation}
Here, we see we can remove the imaginary inertial term and the ac-Stark shift (phase) error with the second quadrature and drive detuning set to
\begin{equation}
		\mathcal{E}^y=-\frac{\mathcal{\dot E}^x}{\Delta} ~\mathrm{and}~\delta_1 =\frac{(\lambda^2-4)\mathcal{E}_x^2}{4\Delta}.
\end{equation} 
Thus, we  eliminate the leakage (second line of \eqrf{DRAG:Ham}) to order ${\mathcal{E}_x^4}/{\Delta^3}$, which in this frame comes about from the induced $0-2$ transition  \footnote{Note this approach is equivalent to a $\mathcal{E}^y$ perturbation using the time-averaged Hamiltonian theory, where $H^V_0=e^{-i\int \sum{\mathcal{E}^y\lambda_j\syo_j}dt}H^xe^{i\int \sum{\mathcal{E}^y\lambda_j\syo_j}dt}$. An adiabatic expansion of $H^V_0$ in terms of $\int\mathcal{E}^y$ is used instead of a Magnus expansion of the unitary operator \cite{Warren:1984a}.}. 
This error can also be removed by adding the term 
 $-i{\lambda\mathcal{E}_x^2}\syo_{0,2}/{8\Delta^2}$ to the transformation $V$. This results in a higher order error being added to the $0-1$ transition, ${3\mathcal{E}_x^3}\sxo_{0,1}/{16\Delta^2}$, which can easily be compensated by rescaling $\mathcal{E}_x$.  Increasing the precision of the transformation up to 5th order results in
\begin{equation}
	\begin{split}
		\mathcal{E}^x(t)=&\mathcal{E}_\pi + \frac{(\lambda^2-4)\mathcal{E}_\pi^3}{8\Delta^2}
		-  \frac{(13 \lambda^4 - 76 \lambda^2 + 112) \mathcal{E}_\pi^5}{128 \Delta^4}, \\		
		\mathcal{E}^y(t)=&-\frac{\mathcal{\dot E}_\pi}{\Delta} + \frac{33(\lambda^2-2)\mathcal{E}_\pi^2\mathcal{\dot E}_\pi}{24\Delta^3} ,\\
		\delta_1(t)=&\frac{(\lambda^2-4)\mathcal{E}_\pi^2}{4\Delta} - \frac{(\lambda^4 - 7 \lambda^2 + 12) \mathcal{E}_\pi^4}{16\Delta^3}.\\
	\end{split}
\end{equation}

\begin{figure}[htbp]
\centering
\includegraphics[width=0.4\textwidth]{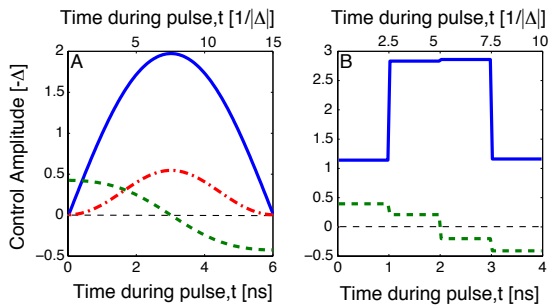}
\caption{ A. 6ns analog Gaussian ($\sigma=t_g/2$) with DRAG; B. 4ns digital GRAPE with no detuning and pixels at 1ns.  In both cases, the blue solid line is the $\mathcal{E}_x$ control and the green dashed is the $\mathcal{E}_y$ control. The red dot-dashed line in A represents explicit detuning, $\delta_1$. Other parameters are the same as in Fig. \ref{fig:PulseWithoutDecay}.}
\label{fig:controls} 
\end{figure}

These pulses remove most of the leakage error, as demonstrated in Fig.  \ref{fig:PulseWithoutDecay}, often with only the first order of corrections. The green dashed (optimized tangential) and red dot-dashed (optimized Gaussian) lines in Fig.  \ref{fig:PulseWithoutDecay} show that DRAG reduces the gate error by orders of magnitude (e.g. 4 orders at 6ns). In fact, we verified with randomized testing that DRAG is largely impervious to envelope shape, provided it has the right area and starts and ends at 0.  Fig. \ref{fig:controls}A shows an optimized Gaussian.

If $\delta_1$ cannot be controlled in real time, phase ramping can be applied \cite{Patt:1992a}. It amounts to transforming to the frame given by transformation  $P=\exp[-i\theta (t) \sum j\Pi_j] $ where $\theta(t) = \int_0^t(\delta_1(s)-\bar\delta_1)ds$ . Then, alternatively, we can pick $\delta_1'=\bar\delta_1=\sfrac{1}{t_g}\int_0^{t_g}\delta_1(t) dt$, 
$\mathcal{E}^x(t)'=\mathcal{E}^x(t)\cos\theta(t)-\mathcal{E}^y(t)\sin\theta(t)$, and
$\mathcal{E}^y(t)'=\mathcal{E}^y(t)\cos\theta(t)+\mathcal{E}^x(t)\sin\theta(t)$.

\section{Numerical optimization (GRAPE)} 
\begin{figure}[htbp]
\centering
\includegraphics[width=0.4\textwidth]{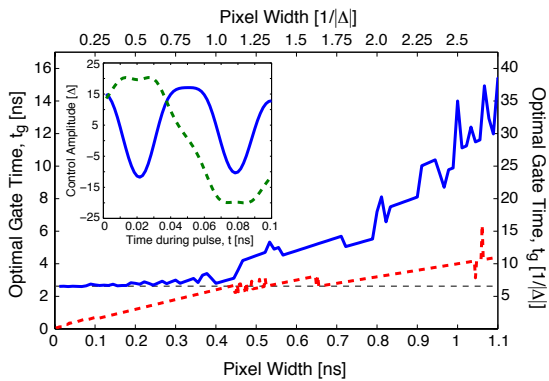}
\caption{Pulse duration vs. pixel width for 1 control (blue solid) and 2 controls (red dashed).  Pulse duration is calculated as the minimal time to bring the error down to $10^{-5}$. The insert shows the two control GRAPE result for $t_g=1/4\Delta$ (blue solid is $\mathcal{E}_x$, and green dashed is $\mathcal{E}_y$). Other parameters are the same as in Fig. \ref{fig:PulseWithoutDecay}.}
\label{fig3}
\end{figure}

We apply optimal control theory to our system by employing the gradient ascent numerical optimization algorithm, GRAPE \cite{Khaneja:2005a}.  We try to minimize the gate error, $1-F_g$, with $F_g$ as defined in Eq. \eqref{eq:fid}.  This is equivalent (for dissipation-free evolution) to maximizing
\begin{equation}
\Phi_2 = \frac{1}{4}(|\bra{0}U_\mathrm{ideal}\dg U_\mathrm{grape}\ket{0}+\bra{1}U_\mathrm{ideal}\dg U_\mathrm{grape}\ket{1}|^2),
\end{equation} 
where $U_\mathrm{grape}$ is the unitary evolution found by GRAPE  \cite{Rebentrost:2008a}. GRAPE works by time-slicing the evolution into discrete time pixels of constant amplitudes $\mathcal{E}^x$ and $\mathcal{E}^y$, as in Fig. \ref{fig:controls}B. The gradient of $\Phi_2$ with respect to each control at each time step provides a  direction in search space.  That is, each iteration of the algorithm increments the controls at each time step of the pulse proportionally to the gradient with respect to the control at that time (including the sign).  This unique search direction allows for fast optimization, particularly with few pixels, and the algorithm quickly converges to an optimal solution ($1-\Phi_2<10^{-5}$) for all gate times
(see Fig. \ref{fig3}).

Restricting GRAPE to 1ns pixels (which is consistent with current experimental limitations \footnote{R. J. Schoelkopf, private communication (2008)}) and setting the initial condition for the $\mathcal{E}^x$ quadrature to be the Gaussian pulse, \eqrf{eq:gaussian}, the algorithm quickly converges to the optimal solution. After optimization, the shape of the Gaussian in the $\mathcal{E}^x$ quadrature is largely unchanged. However the $\mathcal{E}^y$ quadrature changes from 0 to a shape similar to the derivative of the Gaussian, see Fig. \ref{fig:controls}B.  GRAPE essentially takes the idea of DRAG to infinite order and discretizes the pulse. The shapes of the pulses are smooth with no sudden rises or falls in the control amplitudes. The purple dotted line in Fig. \ref{fig:PulseWithoutDecay} plots the gate error as a function of gate time for GRAPE. The extra optimization increases the performance over DRAG.

The two quadrature control GRAPE pulse outperforms its one quadrature control version \cite{Rebentrost:2008a}. Fig. \ref{fig3} shows the minimal pulse duration to obtain gate errors smaller than $10^{-5}$ vs the pixel width for both one (red solid) and two (blue dashed) controls. Here, we see that at small pixel widths (continuous limit) the one control saturates at a time of $2\pi/\Delta$ \cite{Rebentrost:2008a} (black dashed line), whereas the two controls allows for arbitrarily small pulse times. This is because, in the one control case, phase cancelation of the leakage is done through phase accumulation by the natural precession of the third level, whereas, for two controls, it can be done instantaneously. The insert of Fig. \ref{fig3} shows such a continuous two control pulse; this pulse clearly displays a combination of DRAG and composite features \footnote{At this really short gate time the RWA is no longer valid for the Phase \cite{Martinis:2002a,Martinis:2005a,Lucero:2008a} and the Transmon \cite{Koch:2007a,Schreier:2008a,Houck:2008a,Chow:2008a} qubits.}. At the larger pixels (more experimentally realizable), the two controls offer a substantial improvement over one control. For example, at 1ns pixels, there is a factor of 3.5 improvement.

\section{Decoherence} 
\begin{figure}[htbp]
\centering
\includegraphics[width=0.4\textwidth]{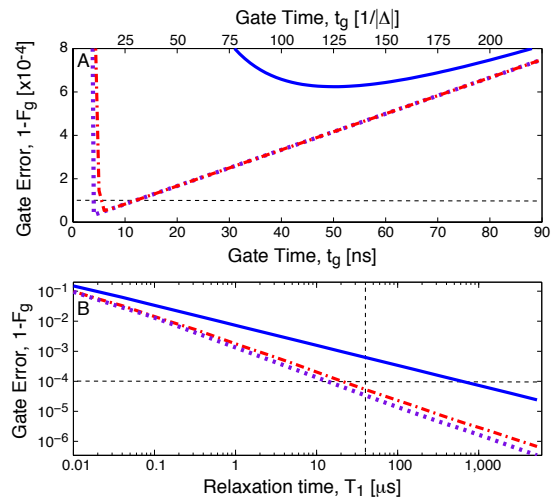}
\caption{A. error vs. pulse duration for $T_1=40\mu$s and B. minimum error vs. $T_1$ for Gaussian ($\sigma$=$t_g/2$) (blue solid), Gaussian ($\sigma$=$t_g/2$) with DRAG (red dot-dashed), and GRAPE (purple dotted) with 1ns pixels. Other parameters are the same as in Fig. \ref{fig:PulseWithoutDecay} and pulses at the optimal times in A are plotted in Fig. \ref{fig:controls}.   }
\label{fig:PulseWithDecay}
\end{figure}

In realistic physical systems, decoherence cannot be neglected. Here, we include this effect by simulating the evolution by a master equation for each optimal pulse. The master equation we use is
\begin{equation}
\dot \rho = - \frac{i}{\hbar} [H^R,\rho] + \sum_{j=1,2}\left[\sfrac{1}{T_1^j}\mathcal{D}[\sigma_j^-]\rho+\sfrac{1}{T_\phi^j}\mathcal{D}[\Pi_j]\rho\right]
\end{equation}
where $\rho$ is the density matrix, $T_1^j$ and $T_\phi^j$ represent relaxation and pure dephasing times respectively, and $\mathcal{D}$ is the damping superoperator, defined as
\begin{equation}
\mathcal{D}[A]\rho=A\rho A\dg - \frac{1}{2}A\dg A\rho - \frac{1}{2}\rho A\dg A.
\end{equation}
For the simulation, we used parameters consistent with the Transmon qubit and as such take pure dephasing to be zero. 
The results are shown in Fig. \ref{fig:PulseWithDecay}. Fig. \ref{fig:PulseWithDecay}A plots the gate error ($1-F_g$) as a function of gate time for $T_1=40\mu$s. Here, we see there is an optimal value for the gate time (when the error due to decoherence is about the same as the error due to leakage) and for both the DRAG (green dashed) and the GRAPE (purple dotted) pulse, this is much less then the standard Gaussian pulse (blue solid) with an optimal error of $5.3 \times 10^{-5}$  and $3.4 \times 10^{-5}$ respectively. Fig. \ref{fig:PulseWithDecay}B plots this minimum as a function of $T_1$ where we observe that at current experimental values $T_1\approx (1-4)~\mu$s \cite{Houck:2008a} DRAG and GRAPE still out preform the simple Gaussian by about an order of magnitude. 

\section{Conclusions} 
The analytical and numerical results presented here demonstrate that it is possible to eliminate leakage in physical qubits by using a second quadrature control approximately equal to the derivative of the first.  The predicted single-qubit error rates are extremely low, even when we include decoherence. While in this paper this procedure is only applied to leakage, it will also find application to other systems, for example turning off unwanted coupling in multi-qubit and qubit-oscillator systems for single and two qubit operations.

\begin{acknowledgments}
We acknowledge W. A. Coish, Jens Koch, B. Osberg, C. A. Ryan, and M. Laforest for valuable discussions. JMG was supported by CIFAR, MITACS, and ORDCF. FM and FKW were supported by NSERC through the discovery grants and QuantumWorks.
\end{acknowledgments}

\end{document}